\pdfoutput=1
\documentclass[fleqn,usenatbib]{mnras}

\usepackage[T1]{fontenc}
\usepackage{ae,aecompl}

\usepackage{hyperref}
\usepackage{graphicx}	
\usepackage{amsmath}	
\usepackage{amssymb}	
\usepackage{epstopdf}

\title[Dark Matter Annihilation Feedback in simulations]{Dark Matter Annihilation Feedback in cosmological simulations I: Code convergence and idealised halos}

\author[N. Iwanus et al.]{
N. Iwanus,$^{1}$\thanks{E-mail: nikolas.iwanus@sydney.edu.au}
P. J. Elahi,$^{2}$
and G. F. Lewis$^{1}$
\\
$^{1}$Sydney Institute for Astronomy, School of Physics, A28, The University of Sydney, NSW 2006, Australia\\
$^{2}$International Centre for Radio Astronomy Research, University of Western Australia, 35 Stirling Highway, Crawley, WA 6009, Australia
}

\date{Accepted XXX. Received YYY; in original form ZZZ}

\pubyear{2017}

\begin{document}
\label{firstpage}
\pagerange{\pageref{firstpage}--\pageref{lastpage}}
\maketitle

\begin{abstract}
We describe and test a novel Dark Matter Annihilation Feedback (DMAF) scheme that has been implemented into the well known cosmological simulation code \textsf{GADGET-2}. In the models considered here, dark matter can undergo self-annihilation/decay into radiation and baryons. These products deposit energy into the surrounding gas particles and then the dark matter/baryon fluid is self-consistently evolved under gravity and hydrodynamics. We present tests of this new feedback implementation in the case of idealised dark matter halos with gas components for a range of halo masses, concentrations and annihilation rates. For some dark matter models, DMAF's ability to evacuate gas is enhanced in lower mass, concentrated halos where the injected energy is comparable to its gravitational binding energy. Therefore, we expect the strongest signs of dark matter annihilation to imprint themselves onto the baryonic structure of concentrated dwarf galaxies through their baryonic fraction and star formation history. Finally we present preliminary results of the first self-consistent DMAF cosmological box simulations showing that the small scale substructure is washed out for large annihilation rates.

\end{abstract}




\section{Introduction}

The Lambda Cold Dark Matter ($\Lambda$CDM) model of the Universe successfully predicts large scale observations, such as the cosmic microwave background \citep{2016Planck2015}, baryonic acoustic oscillations \citep{2005ApJEisenstein, 2016Alam} and supernovae \citep{Riess1998,1999ApJPerlmutter}. However, there appears to be tension when one examines the galactic scales. $\Lambda$CDM N-body simulations predict that dark matter density profiles of galactic halos should appear sharply peaked at their centres \citep{Navarro1996,2001deblok,Oh2008,2010Blok} --- contrary to observed rotation curves measured in nearby dwarf galaxies that imply the cores are flattened out, this is known as the cuspy halo problem \citep{1994NatureMoore}. Milky Way galaxy analogues that are produced in these simulations imply that our galaxy should be isotropically surrounded by many hundreds of dwarf galaxies, whereas observations only show a few dozen \citep{1999ApJWhereAreSatellites} that are aligned preferentially across a large thin disk \citep{2012Pawlowski, 2013Ibata}.

While dark matter and energy are invoked in $\Lambda$CDM, there has been little insight into the fundamental nature of the `Dark Sector'. Dark matter's existence is deduced by its gravitational influence on otherwise discrepant small scale behaviour, such as the rotation curves of galaxies \citep{1980Rubin} and lensing studies of galactic collisions \citep{2004ApJMarkevitch} and leaving the identity of dark matter largely theoretical.

Extensions of the standard model of particle physics for example, yield many potential dark matter candidates \citep{2010Feng}, the most popular class being the Weakly Interacting Massive Particles (WIMPs). WIMPs meet many of the properties needed for a dark matter candidate --- interacting gravitationally, electromagnetically neutral and only coupled to the standard model at or less than weak force coupling. Their popularity compared to other candidates is largely due to theoretical considerations from the early Universe, known as the WIMP miracle. In the WIMP miracle, stable particles are thermally produced when the hot dense plasma from the big bang cools as the universe expands. The abundances of these stable particles are set when the rate of expansion is such that the rate of interaction cannot keep the particles in thermal equilibrium with the rest of the plasma causing these relics to decouple. To reproduce the correct dark matter abundance $\Omega_{\rm DM} \approx 0.258$  \citep{2016Planck2015}, the dark matter particle candidate requires a velocity averaged cross section of $\langle\sigma v\rangle \approx 3 \times10^{-26}$ $\textrm{cm}^{3}$$\textrm{ s}^{-1}$ \citep{2012Steigman}. This result is especially tantalising as this cross section is in the right order of magnitude expected from heavy particles in the GeV/TeV mass ranges that interact through the weak force. Many of these particle have already been postulated for independent theoretical concerns in high energy particle physics \citep{1998Martin, 2005Bertone, 2008Peccei, 2016Gaskins}. As a result most dark matter searches are dominated by the WIMP paradigm.

Direct detection experiments involve measuring the recoil of nucleons colliding with dark matter particles inside large detectors that are built deep underground to minimise environmental radiation noise. Currently the DAMA/LIBRA experiment has a possible detection of dark matter scattering off by an annual modulation in the detector signal-- ostensibly caused by the earth's rotation about the sun \citep{2008Bernabei,2010Bernabei}. Indirect experiments involve the search for dark matter in astrophysical sources by the detection of annihilation products, like gamma rays or electron-positron pairs \citep{2005Bertone, 2008Kuhlen, 2014Adriani, 2014Wechakama, 2016Anderson, 2016Gaskins, 2016Hooper, 2017Liang}. Dwarf spheroidal galaxies in particular have received a lot of interest, as these are regions where the dark matter content is expected to be very dense but with little obscuring dust and gas content \citep{ackermann2014dark}. The Galactic Centre is also another target of interest, serving as a relatively nearby and extremely bright source of gamma rays. Its study requires modelling of diffuse emission through the dust and gas in the centre, but when this background modelling is accounted for in the signal, there is an excess gamma ray emission around 2 GeV which cannot be accounted for. This excess is consistent with a dark matter annihilation through a bottom and anti-bottom quark pairs channel \citep{2015Ackermann, 2016PDUDaylan, 2016Winter}, though an alternate explanation that the excess is due to a population of unresolved millisecond pulsars has recently gained support \citep{Lee2016, Bartels2016, 2016clark, 2016Macias}.

Many searches are model dependent and simplifications have to be made to many aspects of dark matter annihilation, such as the choice of annihilation channels and uncertainty in the dark matter distribution. Because the annihilation rate depends on the square of the density, it is extremely sensitive to the unresolved `substructure' which can provide boosts of many orders of magnitude --- depending on how the measured smooth density field is extrapolated below resolvable scales \citep{ Mack2014}. The presence of unresolvable Ultracompact Minihalos for example can boost the annihilation rate by $10^5$ if they account for even small fractions $(f = 0.01)$ of the total density \citep{2016HCLark2016}. 

Another approach in dark matter research is studying the role dark matter plays in galactic and large scale structure formation. Analytical studies have shown that Dark Matter Annihilation Feedback (DMAF), where the energy from dark matter annihilation is deposited into the nearby baryonic component over cosmic time, can cause the evacuation of gas in smaller halos and inhibit the star formation of the first galaxies \citep{2007Ascasibar, 2008MNRASNatarajan, 2009Natarajan, 2010Ripamonti, 2011Wechakama, schon2015}. The growth of large scale structure is complicated and analytical studies typically rely on linearised perturbations or semi-analytical models which do not capture the non-linear evolution as the density perturbations become large, or the interplay between baryonic and dark matter structure components. This motivates the need to model dark matter annihilation with simulation codes to capture the full effect of of non-linear growth and understand how it impacts the low-redshift Universe -- perhaps imprinting clues from the dark sector. In this paper we present the first attempt at self-consistently following the evolution of annihilating dark matter (which was also used in \cite{2016HCLark2016}, with a boost factor due to Ultracompact minihalo substructure). In Section 2 we outline the equations of our model and outline the modifications to the \textsf{GADGET-2} code. In Section 3 we detail how we produced simple test halos and test our code for different halos sizes, annihilation rates and check for convergence. In Section 4 we give an overview of how we might expect DMAF to affect the universe and present preliminary results from cosmological box simulations.

\begin{figure}
	\includegraphics[width=\columnwidth]{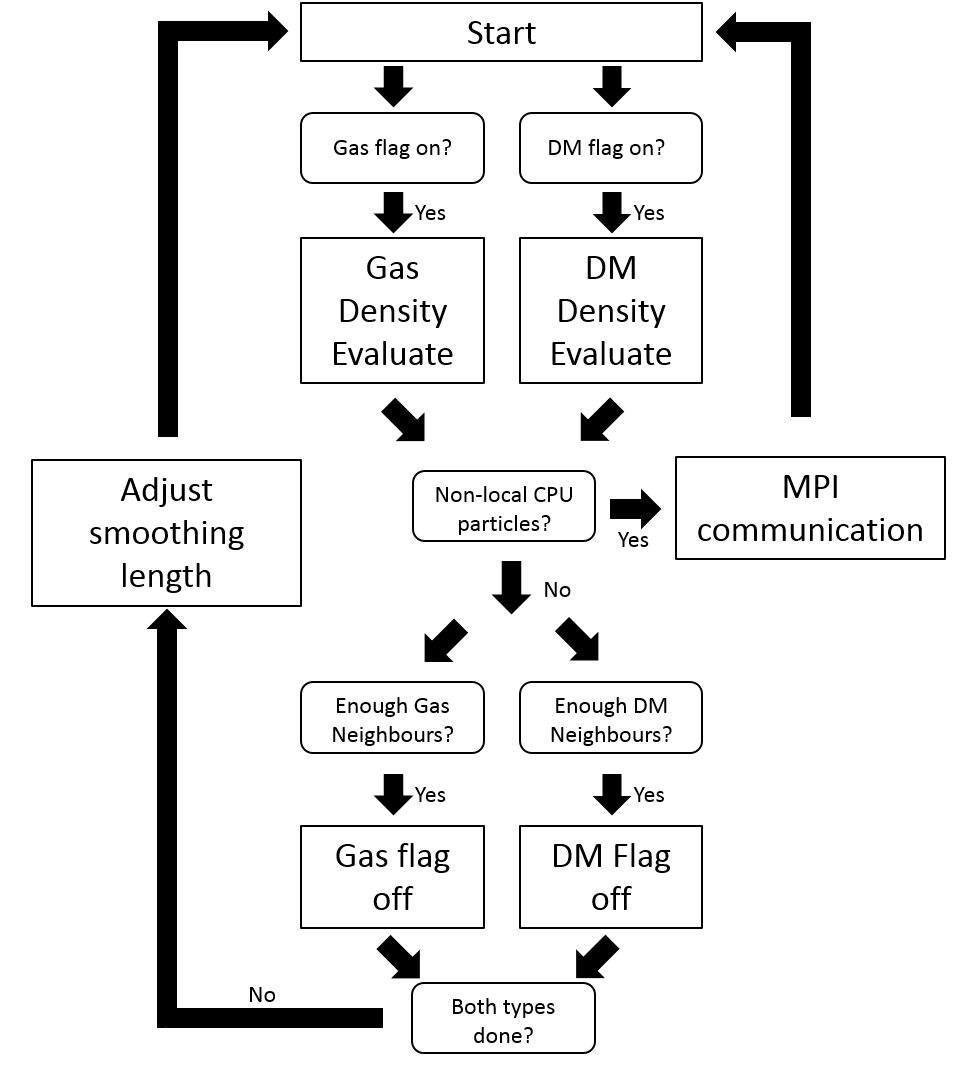}
    \caption{The modified density loop flow chart. Other smaller differences include altering \textsf{GADGET-2} data structures so that SPH particles hold dark matter properties (e.g dark matter smoothing lengths and density), inclusion of a dark matter nearest neighbour search function and DMAF energy injection in the second SPH loop.}
    \label{fig:density_loop}
\end{figure}

\section{Annihilation Feedback}
We modified \textsf{GADGET-2} \citep{Springel2005a}, a parallel N-body simulation code, implementing a dark matter annihilation/decay feedback scheme. \textsf{GADGET-2} and its descendants are ubiquitously used in the cosmological community for studies of structure formation. Gravitational forces are calculated using the TreePM method by a set of N-body particles. The hydrodynamical forces in the code are calculated utilising the Smooth Particle Hydrodynamics (SPH; \citeauthor{Monaghan1977SPH} \citeyear{Monaghan1977SPH}) formulation of fluid dynamics, where fluid elements are represented by discrete particles. In the traditional SPH formulation\footnote{There have been numerous advances since traditional SPH that improve its treatment of instabilities and shocks \citep{2010Read, 2010Springela, 2013Hopkinsa, 2015Hopkinsa, 2016Sembolini}. However, although we use classic SPH here, the scheme is not reliant on a particular SPH implementation or even limited to SPH in principle. It could also be used in mesh and moving mesh codes, estimating the density field at cell centres.}, the density of a particle $i$ is approximated by

\begin{equation}
    \rho_i = \sum_{j=1}^{N}  M_j W(r_{ij},h_i),
	\label{eq:sph_density_estimate}
\end{equation}
where the sum is over nearby gas particles labelled $j$, mass of the particle is $M_j$ weighted by a kernel $W(r,h)$ that is a decreasing function of the distance between particle pairs $r$ and a smoothing length $h$. Typically $h$ is chosen such that there are $N$ nearest neighbours, or that a constant mass is held within a kernel that falls to zero at $h$. We use the standard B-spline kernel, though higher order kernels could be used.

\subsection{Annihilation Implementation}
\label{sec:maths} 
We assume a constant velocity-averaged annihilation cross section $\langle \sigma v \rangle$ for a two-body interaction rate per volume that is proportional to the number density of dark matter squared. Writing the annihilation rate in terms of the physical dark matter density $\rho^{2}_{\chi}$ we get:

\begin{equation}
    \frac{dA}{dt} = \frac{\langle \sigma v \rangle}{2m^{2}_{\chi}} \rho^{2}_{\chi},
	\label{eq:annihilation_rate}
\end{equation}
where $m_{\chi}$ is the dark matter particle mass. To implement this rate into the code, we use the SPH estimate of $\rho_{\chi}$, (equation \ref{eq:sph_density_estimate}), where now we sum over nearby DM particles instead of gas. From Equation \ref{eq:annihilation_rate}, we can calculate the quantities of interest -- the dark matter mass annihilating away as well as the power released into nearby gas. The mass loss rate is given by

\begin{equation} \label{eq:masschange}
\begin{split}
\frac{dM_i}{dt} & = -\frac{\langle\sigma v \rangle}{m_{\chi}}\rho_{\chi} M_{i},
\end{split}
\end{equation}
where in each annihilation $2m_{\chi}$ units of mass are lost and we have estimated that the volume occupied by a DM particle is $V = M_{i}/\rho_{\chi}$.

Annihilation products, like electron-positron pairs also couple to the gas and deposit their energy, the so-called Dark Matter Annihilation Feedback. Again we use equation (\ref{eq:annihilation_rate}) to estimate the annihilation rate --- now at the position of each gas particle. The energy produced at each gas particle is equal to $2m_{\chi}c^{2}$. For this work we assume that the mean deposition length scale of energy is within SPH smoothing lengths, so all heating is local and perfectly couples with the gas. With these assumptions the energy deposition rate is

\begin{equation}
   \frac{du_{i}}{dt} = \frac{\epsilon\langle\sigma v \rangle c^{2}}{m_{\chi}}\frac{\rho_{\chi}^{2}}{\rho_{gas}}.
	\label{eq:gas_injection}
\end{equation}
We stress that $\rho_{\chi}$ is an estimate of the local DM density using equation (\ref{eq:sph_density_estimate}) summing over the nearest DM particles, while $\rho_{gas}$ is the same equation but now summing over the N nearest gas neighbours and $\epsilon$ is the coupling efficiency, were we have assumed $\epsilon = 1$, though it could in principle be less due to a neutrino decay channel for example. This method is computationally efficient, as it requires only one extra density calculation loop per particle. In addition our method can be trivially adapted to the case of decaying dark matter by simply changing the code to calculate a rate linearly proportional to the local dark matter density and adjusting the appropriate constants.

A weakness of this scheme is that as the mass loss equation (\ref{eq:masschange}) and energy injection rate equation (\ref{eq:gas_injection}) are sampled over the DM particles and gas particles respectively, they measure the DM density fields at different positions (separations $\approx h$), hence the total amount of energy transferred between the two components will only be approximately conserved. 

However, a formally rigorous treatment requires breaking up annihilation products into short and long range energy deposition, tracing the products from their origin dark matter particles. Given the uncertainty of what their annihilation channels are and their coupling efficiencies, local energy deposition is a reasonable assumption.

\subsection{Code implementation}

\begin{figure}
	\includegraphics[width=\columnwidth]{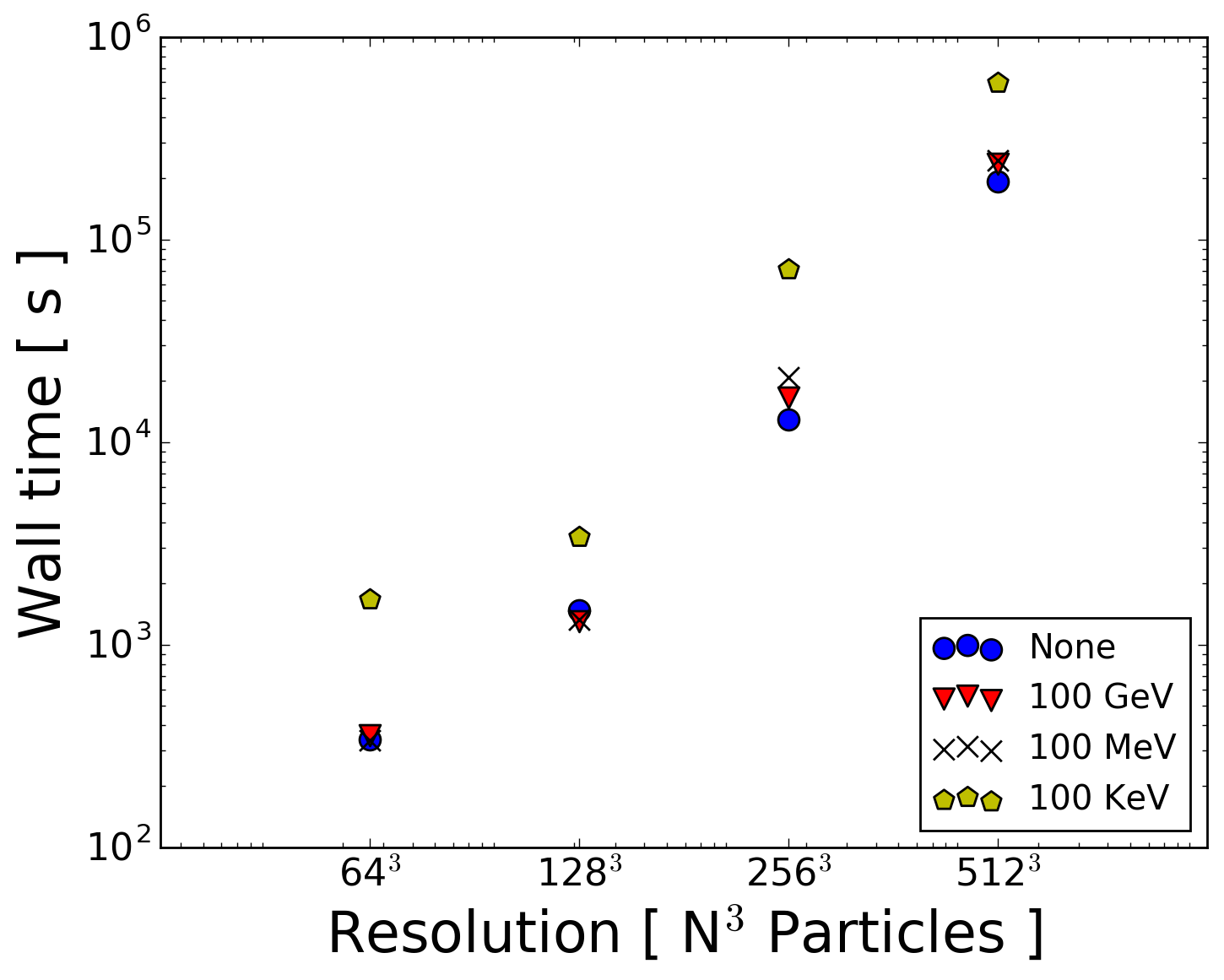}
    \caption{The wall-time taken to run a DMAF 100 Mpc box simulation with different resolutions. Initial conditions are identical and generated with \textsf{N-Genic}. These simulations were run on \textsf{ARTEMIS} High Performance Supercomputer at the University of Sydney. The runs used 120 cores with identical simulation parameters, only varying the mass of the dark matter particle m$_{\chi}$ and softening lengths set at 5 percent of the initial inter-particle spacing. The 100 keV simulations are most adversely affected due to having the largest amount of energy injection. }
    \label{fig:cpu_runtime}
\end{figure}

\begin{figure*}
	\includegraphics[width=2\columnwidth]{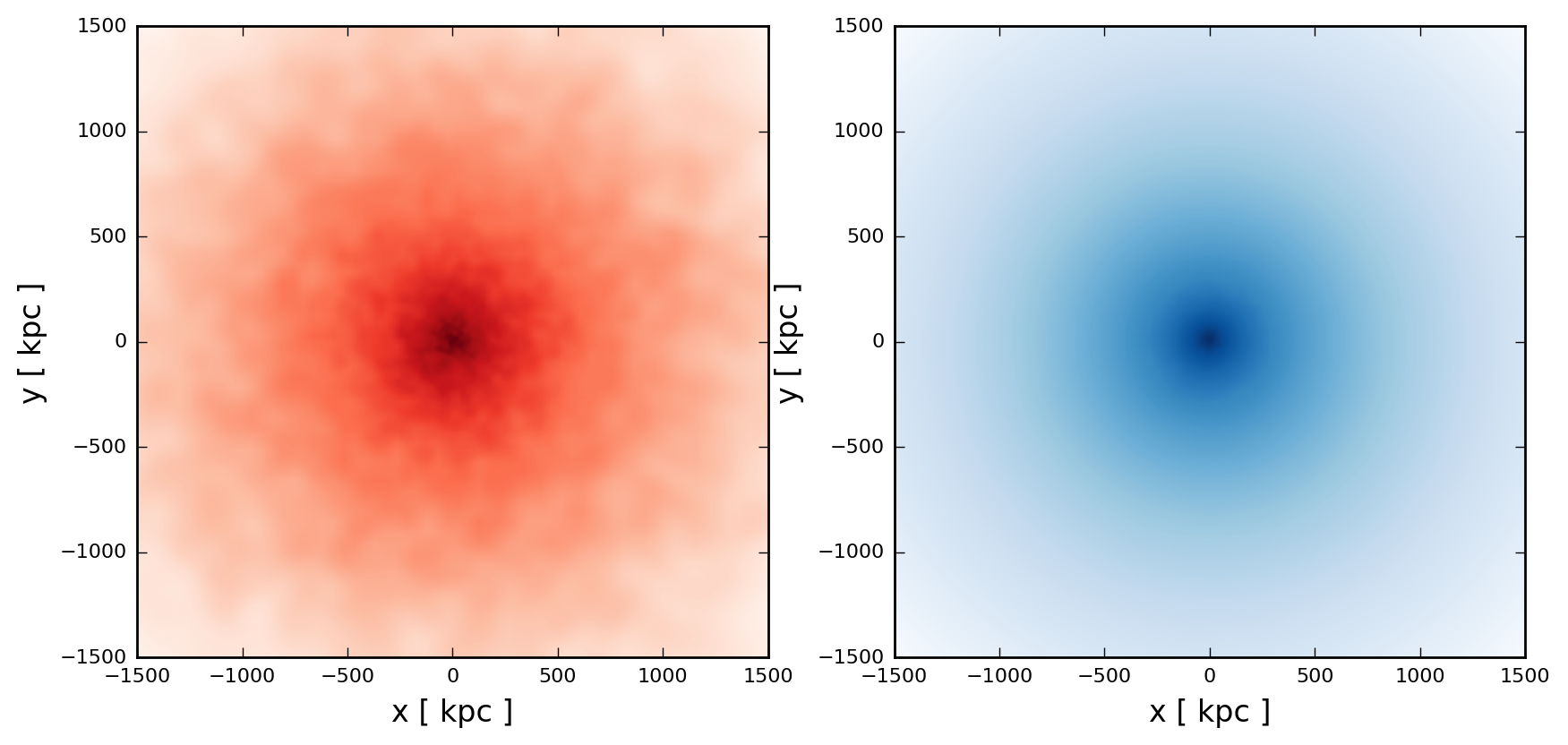}
    \caption{An example of one of the halos used in our tests showing the dark matter (left) and gas (right) density of a large NFW profiled halo, $(M = 10^{15}$ M$_{\sun}$, $c$=6.9 ) sampled with 10$^5$ particles. Halos like this were created using \textsf{GalactICs} and would serve as our initial conditions for the purpose of testing our codes. The gaseous component was created by converting approximately $\Omega_{\rm b}/\Omega_{\rm CDM} \approx 20\%$ of the dark matter particles into gas and assigning a thermal energy equal to its local velocity dispersion. These ICs were allowed to relax and re-virialize over 20 Gyrs. The intensity corresponds to the SPH density integrated down the z-axis and was created with the package \textsf{Pynbody} \citep{pynbody}.}
    \label{fig:SPH_DM_GAS}
\end{figure*}

The main changes to the code involve the creation of new members to \textsf{GADGET-2}'s particle and MPI communication structures and modifications to the first density loop, see figure \ref{fig:density_loop}. In order to calculate equation \ref{eq:gas_injection} we modify the density loop to search for the nearest dark matter neighbours of the gas particles. We then alter the subsequent MPI communication process to send the relevant variables for the dark matter density calculations in addition to the regular baryonic variables. The termination condition of the loop for each particle is changed such that the number of gas and also its dark matter neighbours must be within the specified tolerance, otherwise the dark matter and/or gas smoothing lengths are adjusted and the density loop repeats. To avoid unnecessary recalculation, we tag each particle with a gas and dark matter flag, signalling that we have the correct number of SPH neighbours for that particle type and skip the corresponding density calculations. The other parameters are now just constants of the specific dark matter model and the energy injection itself occurs at the end of the second SPH loop after the entropy generation rate by shock heating is calculated to which Equation \ref{eq:gas_injection} is added.

The extra CPU time cost of these routines is expected to mostly come from the extra nearest neighbour searches and MPI communication. Naively, now that  $\rho_{gas}$ and $\rho_{\chi}$ are calculated for each gas particle, we should expect that the CPU time within the hydrodynamics part of the code is increased by a factor of 2, assuming the nearest neighbour search is roughly equal for gas and dark matter neighbours. In practice however the SPH smoothing lengths are set independently, requiring particles to loop until both the numbers of dark matter and gas neighbours are within the set tolerance. The number of MPI communications that are expected to occur between threads will increase as well, since there exists the possibility that a dark matter particle will need to be communicated even if the gas loop is local or complete. So overall we expect the CPU time increase to be some constant factor slightly greater than 2 in the density loop. Practical tests of adiabatic cosmological box simulations and DMAF (results of which we will present in a future paper) seem to indicate an overall time cost, see figure \ref{fig:cpu_runtime}, of a factor of less than 2 for modest annihilation rates in large N particle simulations. In the runs 100 keV, the performance is much worse due to the large amount of energy injected. The increased pressure and subsequent large hydro-forces activate \textsf{GADGET-2}s time adaptive integration scheme, forcing smaller steps to maintain accuracy of the code and a larger amount of  time spent searching for neighbours.

Despite good performance, the code will be further optimised in the future. A reduction in  MPI communication time can be achieved by reducing transfer of unneeded data structures during processor communication i.e sending only dark matter properties instead of gas and dark matter properties, could lead to a small speed up but is expected to be minimal as most of increase is due to longer searches. Currently the nearest neighbour search is conducted by `walking' \textsf{GADGET-2}'s oct tree data structure \citep{1986Barnes} and picking up particles of the correct type within the smoothing length. A larger speed up is expected if we streamline the search-- for each walk picking up both dark matter and gas particles within the smoothing length instead of two separate searches.

\subsection{Code Verification on simple boxes and  galaxies}\label{sec:verification} 
We have tested our code in simple uniform boxes, static and expanding and compare to analytical solutions for the overall heating. The tests clearly show errors in the energy deposition of < 2 percent even in extreme models. This error can be significantly reduced to <0.3 percent by using smaller time-steps. The remaining error we associate with deviation in the SPH reconstruction of the uniform density and we find we can further improve this with a greater particle resolution, see Appendix (\ref{box_solutions}). The error associated with the larger time-steps imply the need for a limiter as in \cite{2009Saitoh} to ensure stability in the the heating rate, particularly for our cosmological runs. Nonetheless we have checked over our simulations here to ensure no shocks associated with this time-step instability developed.

We also tested our code in the simple case of ideal, isolated halos which formed the backbone of our tests. They were modelled them using the Navarro-Frenk-White (NFW) profile \citep{Navarro1996} given by :

\begin{equation}
    \rho (r)=\frac{\rho_0}{ (\frac{r}{a})(1 + \frac{r}{a})^2},
	\label{eq:nfw}
\end{equation}
where $a$ is the scale radius and $\rho_{0}$ a normalising factor that determines the mass. We used \textsf{GalactICs} \citep{galactics1, galactics2, galactics3}, to produce our test halos. For dark matter only simulations we evolved the halo for 20 Gyrs to allow initial instabilities in the halo to relax. For dark matter annihilation feedback simulations, $\Omega_{\rm b}/\Omega_{\rm CDM} \approx 20\%$ of the dark matter particles were converted into gas with the same mass, a thermal energy equal to their local velocity dispersions and the kinetic energy correspondingly reduced to account for this additional thermal energy. These simulations were allowed to relax over 20 Gyrs, though a small amount of mass was initially shed due an instability caused by the gravitational force softening, not accounted for in \textsf{GalactICs} at the central cusp (typically only a few percent of the mass was lost). The end results are stable halos, see figure \ref{fig:SPH_DM_GAS} and these would serve as our initial conditions (ICs) for testing the code.

\section{Results}

\subsection{Annihilation mass reduction}\label{mass_loss_results}
\begin{figure}
	\includegraphics[width=\columnwidth]{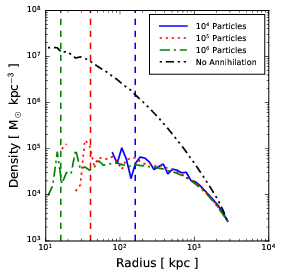}
    \caption{To test the convergence of our mass loss implementation (equation \ref{eq:masschange}), we ran 3 simulations of an NFW halo ($M = 10^{15} M_{\sun}$, $c=6.9 $) with different particle number resolutions and an extreme annihilation model with mass loss due to an 0.1 eV thermal relic dark matter particle, run for 20 Gyrs. We have good agreement everywhere except towards the centres of the halo where the smaller sampling volume leads to noise in the inner regions. The vertical lines show the gravitational softening length used for the corresponding simulation.}
    \label{fig:convergence}
\end{figure}

For the dark matter only simulations, we studied a halo of mass  $M= 10^{15} M_{\odot}$ and concentration 6.9 and varied the number of particles to show that in our smooth NFW profiled test halo, the simulation results converge. In figure \ref{fig:convergence} we plot the evolved dark matter density profiles of the halo after 20 Gyrs and DMAF from an extreme model, a particle with mass of 0.1 eV, for the same halo but sampled with a different number of particles. The profiles nicely converge down to 100 kpc for our lowest resolution simulation. Further in, the halo is not well sampled and noise occurs at a radius that is near the gravitational smoothing lengths of the simulations.

\begin{figure}
	\includegraphics[width=\columnwidth]{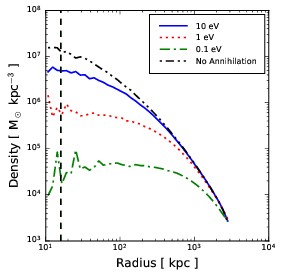}
    \caption{A pure dark matter simulation of an NFW profiled halo ($M = 10^{15} M_{\sun}$, $c=6.9 $), was evolved for 20 Gyrs. The simulation with the lowest dark matter mass (highest rate) exhibit a suppression of the density at the centre of the halo where the annihilation is strongest.}
    \label{fig:profile}
\end{figure}

In figure \ref{fig:profile}, we again take the same ICs sampled by $10^6$ particles and evolved 20 Gyrs but now vary the particle mass. We see that the loss of mass is most prominent in towards the centre of our halos, as expected in the high density regions. The models used in these runs are quite extreme for the purposes of highlighting the two cases with and without DMA, the highest mass particle being 10 eV annihilating with the thermal relic cross section. Despite this, the total mass difference during this time period only amounts to a total mass change of 2 percent for the highest rate. Unless the annihilation cross section (or dark matter particle mass) is much higher (lower) than what observational data permits, it seems that the overall direct change of mass due to dark matter annihilation is negligible. 

\subsection{Gas with DMAF}

\begin{figure*}
	\includegraphics[width=2\columnwidth]{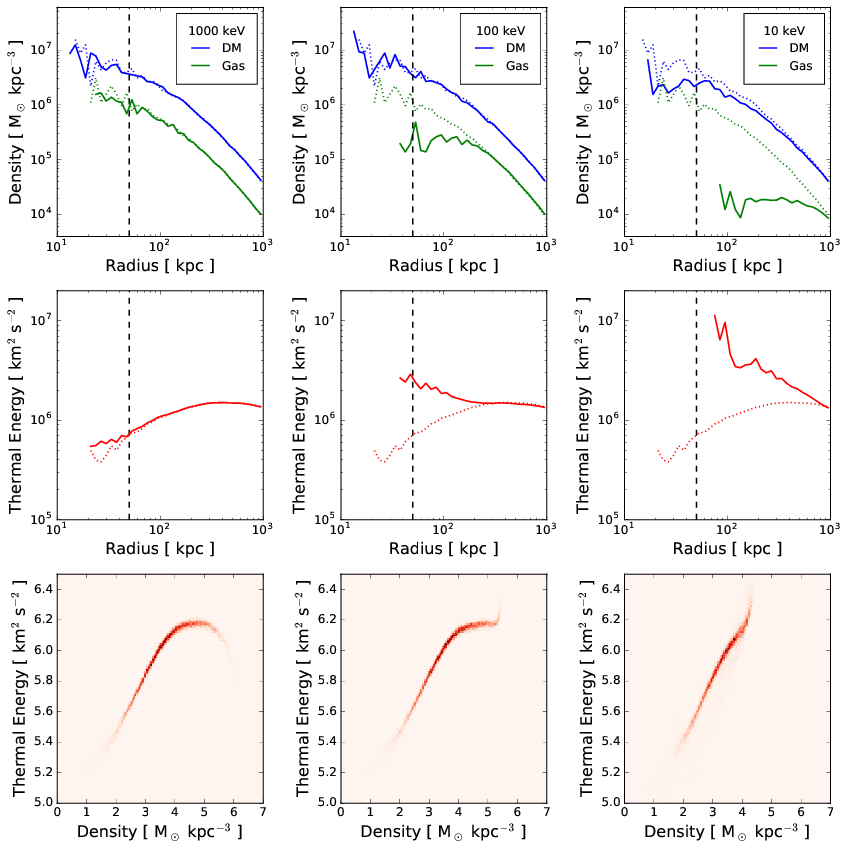}
    \caption{DMAF simulation results under different particle mass models. The cross section was set at $\langle\sigma v\rangle = 3\times10^{-26}\textrm{cm}^{3} \textrm{s}^{-1}$ and DMAF energy is deposited into an NFW profiled halo ($M = 10^{15} M_{\sun}$, $c=6.9 $), for 10 Gyrs. The top row of panels show the change in the dark matter and gas profiles due to DMAF (solid lines) compared with non-DMAF simulations (dotted lines) and the vertical dashed lines are the gravitational smoothing lengths. The middle row panels likewise compare the thermal energy profile evolved with and without DMAF. The bottom rows shows the density-thermal energy distribution of the DMAF simulation, showing the high density gas is preferentially heated due to the halo cusp.}
    \label{fig:nfw_profile_halos}
\end{figure*}

We ran simulations for a halo with similar properties but now with 20 percent of the dark matter particles randomly converted to gas and stabilised after 20 Gyr. These halos were then used as the initial conditions for simulations with DMAF energy injection given by equation \ref{eq:gas_injection}, that ran for 10 Gyrs, see figure \ref{fig:nfw_profile_halos}. DMAF preferentially deposits a large amount of thermal energy into the gas particles located within the high DM density regions at the centre. The resultant pressure gradients evacuate large amounts of gas; in the inner regions of the halo the radial density decreases by 2-3 orders of magnitude for our strongest DMAF simulation (10 keV).

\begin{figure*}
	\includegraphics[width=2\columnwidth]{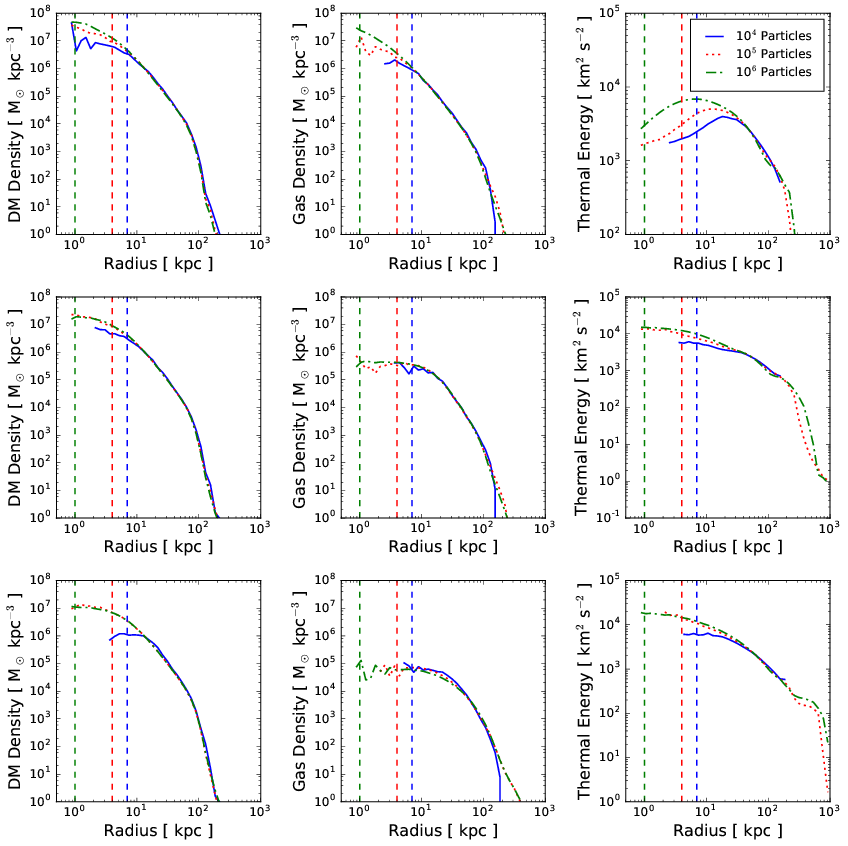}
    \caption{Initial conditions for a convergence run (top row) and the DMAF simulations after 10 Gyrs (mid row panels) and 25 Gyrs (bottom row) of a Milky Way sized halo $(M = 10^{12}$ M$_{\odot} , c= 15)$ sampled at three different resolutions and undergoing DMAF from a thermal relic 100 MeV dark matter particle. The dark matter density profiles (left) and the gas density profiles (centre) are initially near identical except for the inner regions near the smoothing length. The thermal energy (right panel) was set by each particles local velocity dispersion and adjusting the kinetic energy appropriately (see section \ref{sec:verification}). Though the thermal energy profile differ initially in the inner regions and far outer edges, we find that with evolution under DMAF, the differences shrink and converge as the injected energy becomes the dominant component in the gas.}
    \label{fig:convergence_gas}
\end{figure*}

Having shown the code depositing energy into the gas as expected, we now turn to the question of convergence. We ran DMAF simulations on a Milky way sized halo $(M = 10^{12} M_{\odot} , c=15)$ sampled at 3 different resolutions, the initial conditions are shown in the first row of figure \ref{fig:convergence_gas}. Although initially the thermal energy profiles differ, we find good convergence of the results with increasing resolution in both the dark matter and gas profiles over large simulation times. For the highest particle resolution runs, over the course of 25 Gyrs, the differences in the thermal profiles wash away as the annihilation energy is deposited. In the lowest resolution simulation, near the gravitationally softened region, the dark matter profile is relatively flatter than its higher resolution counter parts. As a consequence the energy injected into the halo is correspondingly lower. In the higher resolution simulations however, the dark matter profiles more closely agree and thus so do the thermal energy profiles. This demonstrates the dependence our scheme has on the resolution of the dark matter resolution, the thermal energy injected converges when the dark matter distribution is also converged.

\begin{figure*}
	\includegraphics[width=2\columnwidth]{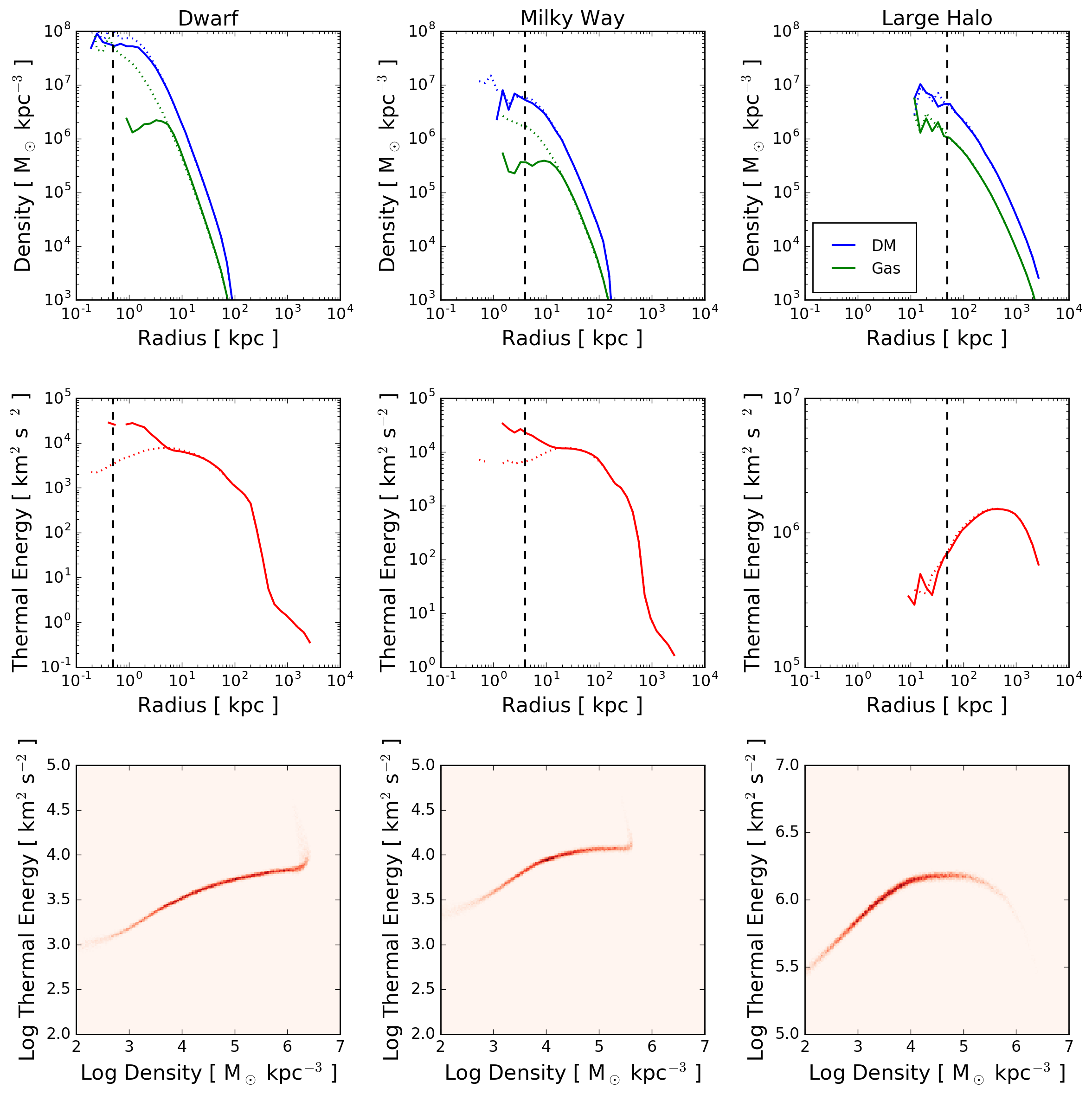}
    \caption{A comparison of the effect of DMAF on three different types of halos. From left to right we show a dwarf sized halo $(M = 10^{10}$ M$_{\odot} , c= 25)$, a Milky way sized halo $(M = 10^{12}$ M$_{\odot} , c= 15)$ and a large halo $(M = 10^{15}$ M$_{\odot} , c= 6.9)$. These panels show the evolution after only 0.5 Gyrs of evolution under DMAF from a 100 MeV dark matter particle (dotted lines show simulations without DMAF).}
    \label{fig:different_halos}
\end{figure*}

DMAF should influence galaxies most when the amount of energy injected into the halo is similar to its gravitational binding energy. Indeed we see galaxies with less mass more readily eject their gas, see figure \ref{fig:different_halos}. These halos of different masses and concentrations but with a identical DMAF parameters, 100 MeV and cross section of $\langle\sigma v\rangle = 3\times10^{-26} \textrm{cm}^{3}  \textrm{s}^{-1}$ all evolved for only 0.5 Gyrs and we see that while the largest of the halos is barely affected, the smaller but concentrated ones are altered substantially. In the `dwarf halo' case, the effect is drastic enough such that we see that even the DM profile is affected at radii greater than the simulation smoothing length.

\begin{figure*}
	\includegraphics[width=2\columnwidth]{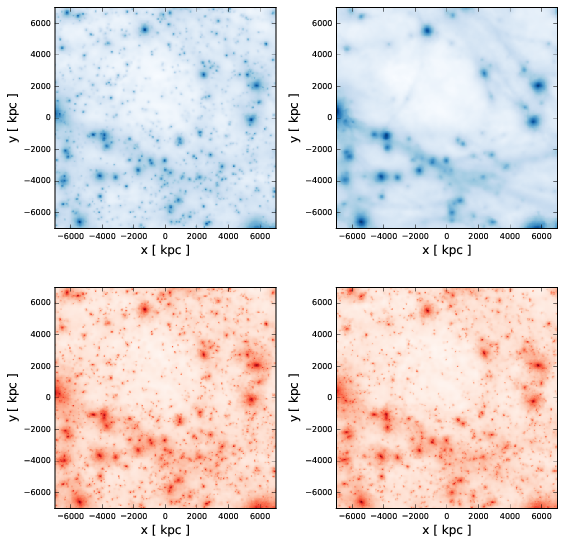}
    \caption{SPH images of box simulations (dark matter is shown in red, gas in blue) of a $\Lambda$CDM universe  (left hand panels)  and the corresponding simulations for a universe with DMAF (right panels) from a 100 MeV dark matter particle annihilating with the thermal relic cross section. Due to annihilation from this model, the gas is blown out from most galaxies except for the largest. Because of the loss of the gravitational pull by the ejected gas many of the smaller dark matter structures have also been washed away. The intensity corresponds to the SPH density integrated down the z-axis and was created with the package \textsf{Pynbody} \citep{pynbody}.}
    \label{fig:box_sims}
\end{figure*}

\section{Annihilation feedback in a cosmological context}

Dark matter annihilation simulations have been performed by \cite{2012ApJRJSmith} where they applied a model of DMAF in the context of protostars forming around dark matter minihalos. They found that for modest annihilation rates (thermal relic cross sections with 100 GeV mass), DMAF prevents fragmentation or at least preferential formation of wide binary stars subsequent to collapse. However their treatment did not simulate a `live' dark matter halo and instead used a spherically symmetric background halo and accreting gas was then simulated on top of this analytical halo on very small scales ($ \sim 0.01$ pc). \cite{2008MNRASNatarajan} studied DMAF in a large-scale cosmological context. Their treatment involved semi analytical models of baryon collapse \citep{2006MNRASCroton} again with analytical halos albeit with NFW parameters fitted from the Millennium dark matter simulation runs \citep{2005Springel}. Nonetheless, \cite{2008MNRASNatarajan} found that in very sharply sloped halos, the annihilation rate could be large enough to offset the galactic cooling rate and have an appreciable effect on the gas distribution and luminosity function of galaxies. 

A main advantage of using analytical halos instead of `live' simulations is that it sidesteps issues of resolving the high density peaks towards the centres of the halos --- which would otherwise underestimate the total annihilation emanating from  within an unresolved cusp. This comes at the cost of not having a `true' dynamic halo that is interacting with gas. It has been shown that in full N-body simulations, baryonic feedback events like supernovae and star formation can cause rapid gravitational instabilities that facilitate the formation of flat cored profiles \citep{1996MNRAScoresNavarro, 2010ApJlackner, 2012MNRASPontzen}. It is therefore important to not only consider dark matter annihilation feedback on the gas but to study it in the presence of `live' halos with complicated formation histories, substructure and mergers. To this end we have the first cosmological box simulations to create a sample of galaxies and study DMAF on a more realistic population.

Though our models have `unresolved cusps' smoothed over by gravitational softening, their size is simply tied to scale of how well the dark matter density is resolved. In the case of our halos, this is given by the power criterion \citep{2003Power} i.e resolved down to near the gravitational softening length. From this it follows that the correct amount of DMAF energy will only be deposited when the size of a halo's physical core is larger then the smoothing length. But even in the case of an infinitely `cuspy halo' with no core, while the energy density injection (\ref{eq:gas_injection}) in principle diverges, the total energy injected into the halo does not. DMAF injection scales with $\rho _{\chi}^{2} \propto r^{-2}$ in our halos, while the volumes at which these energies are deposited at scale with with $V \propto r^3$. Hence we should find the relative uncertainty in energy injected will be at most the ratio of the unresolved core size to the scale radius of our halos, see Appendix (\ref{App:DMAF_halo_error}).

While our idealised halo simulations lack more complicated physics, such as realistic cooling curves, they show that annihilation increases in importance for less massive but concentrated halos, due to weaker gravitational binding and highly concentrated density peaks.  Incorporation of DMAF into a more complete code with the advanced physics models (cooling, stars, etc) as well as a more realistic model for annihilation products coupling to gas is in the works. Furthermore our assumption of 100 percent  heating efficiency is only approximately true for light dark matter models at high redshifts \citep{2007_IGMHEAT_Ripamonti, SPF, 2012Evoli2012}. DMAF coupling on galactic scales is an ongoing area of research with large uncertainties. Nonetheless our interpretations are consistent with \cite{schon2015} who found that DMAF in the form of electron-positron annihilation pairs couple efficiently to the gas --- enough such that DMAF is likely to have an effect on primordial halos of mass $10^5 - 10^6 M_{\odot}$ at high red-shifts $z>20$, even for heavy dark matter models (100 GeV) which are far less constrained by observations. In the context of the Milky Way cosmic ray electrons, \citep{2010Delahaye} find energy deposition lengths of order $\approx 1   \rm{kpc}$.

Halos with present-day masses of about $10^{10} M_{\odot}$ are especially interesting, as below this mass scale the baryon fraction of these dwarf halos are significantly reduced compared to cosmological abundances \citep{Papastergi2012}. This is largely due to UV-suppression feedback \citep{2006MHoeft} balancing the gas cooling rate and preventing its condensation. Additional heating sources, like unaccounted DMAF which precisely affects these smaller, concentrated halos more prominently, as we saw in figure \ref{fig:different_halos}, could therefore tip the balance in this regime pushing up the baryon fraction drop off scale. Therefore dwarf galaxies in particular will be a focus of our cosmological runs. Indeed preliminary results from our cosmological runs back this assertion up. Figure \ref{fig:box_sims} shows gas (blue) and dark matter (red) integrated density images of a subsection of our box in the case of plain $\Lambda$CDM (left panels) and with DMAF from a 100 MeV dark matter particle (right panels). The halos appear much less dense and much of the small galactic structure has been washed away. More so, while the effect is not as striking as in the gas image, many of the smaller dark matter halos have been washed away as well. A fuller analysis of these simulations will be made available in the next paper of this series.

\section{Conclusions}
We present the first hydrodynamical simulations to self-consistently model Dark Matter Annihilation Feedback. New routines are implemented into the simulation code \textsf{GADGET-2} that take into account the effect of dark matter annihilation products depositing energy into the nearby baryonic structure of galaxies. We tested this code  on uniform box simulations and  the case of simple galaxies modelled as NFW halos with a gas component ($\Omega_{\rm b}/\Omega_{\rm CDM} \approx 20\%$). As expected DMAF preferentially heats the gas at the centre of the halo near the dark matter density peak. Subsequently the heated gas rises out of the halos altering their density and thermal profiles. We repeated our simulations at different resolutions to confirm convergence of our results and found good agreement between realisations of similar halos sampled with $10^4, 10^5$ and $10^6$ particles. Testing on halos with different masses and concentrations show that DMAF is most likely to affect the smaller scale structures of galaxies such as sub halos and the dwarfs through their star formation histories, X-ray profiles and baryonic content. Preliminary results of dark matter and gas cosmological box simulations (100 Mpc/h, 512$^{3}$ Particles), with DMAF from 100 MeV dark matter particles, show much of the baryonic small scale structure is washed away-- leaving only the largest gaseous structures and even influencing small dark matter structure.  Much effort has been invested by simulators into generating hydro simulations and feedback prescriptions that reproduce small galaxies \citep{2016MNRASTollet, 2016Sawala, 2016MNRASVogelsberger, 2016Read, 2016MNRASChen, 2016Fitts, 2017GarrisonKimmel}. Our results indicate that DMAF must be considered when modelling this regime and simulations with sufficient annihilation rates could have important implications on the `cuspy halos' and `missing dwarfs' problems in cosmology.

\section*{Acknowledgements}

The authors thank Chris Power for his helpful feedback.  The authors also acknowledge the University of Sydney HPC service for providing computational resources from \textsf{ARTEMIS} which has contributed to the research results reported within this paper as well as the assistance of resources and services from the National Computational Infrastructure (NCI), which is supported by the Australian Government. Nikolas Iwanus is supported by an Australian Postgraduate Award (APA).




\bibliographystyle{mnras}
\bibliography{paper1bib3} 




\appendix
\section{DMAF in Uniform box solutions}
\label{box_solutions}

In a homogeneous and isotropic universe, the densities are constant with respect to position and so equation \ref{eq:gas_injection} can be written as
\begin{equation}
   \frac{du}{dt} = \frac{\langle\sigma v \rangle c^{2}}{m_{\chi}}\frac{\rho_{crit}}{a(t)^3}\frac{\Omega_{\chi}^{2}}{\Omega_{b}},
	\label{eq:gas_injection2}
\end{equation}
where $\Omega_b$ and $\Omega_{\chi}$ are the usual baryon and dark matter cosmological density parameters, $\rho_{crit}$ is the critical density and $a(t)$ is the scale factor as determined by the Friedman equations. In the case of a static box $a(t)=1$, the solution simply integrates a constant and so the thermal energy increases linearly. For an expanding box we transform from $t$ and integrate in $a(t)$ space. For the case of a flat matter dominated universe ($\Omega_{M} = 1$, $\Omega_{b} = 0.1573$,  $\Omega_{\chi} = \Omega_{M} - \Omega_{b}$), the expanding Friedman equation reads

\begin{equation}
   \frac{da}{dt} = H_{0}a^{-\frac{1}{2}}
	\label{eq:Flat_matter_dominated_friedman},
\end{equation}
where $H_0$ is the Hubble constant. Dividing equations \ref{eq:gas_injection2} by \ref{eq:Flat_matter_dominated_friedman} and adding a term $-\frac{u}{a}$ to take into account cosmological adiabatic expansion, yields the heating rate

\begin{equation}
      \frac{du}{da} = \frac{\langle\sigma v \rangle c^{2}}{m_{\chi}}\frac{\rho_{crit}}{ H_{0} a^{\frac{5}{2}}}\frac{\Omega_{\chi}^{2}}{\Omega_{b}} -\frac{u}{a}.
	\label{eq:Flat_matter_dominated_friedman_Step}
\end{equation}
Collecting the constants of the first term on the RHS into $\kappa$ and using an integrating factor yields the solution

\begin{equation}
   u(z) = \frac{1+z}{1+z_0}\Bigg[ u_0  + 2\kappa(1+z_0)^{\frac{3}{2}}\Big(1- \Big(\frac{1+z}{1+z_0}\Big)^{\frac{1}{2}}\Big) \Bigg]
	\label{eq:Flat_matter_dominated_friedman_Final_equation},
\end{equation}
where z is the redshift and we have used $a=\frac{1}{1+z}$, $u_0$ is any initial thermal energy at $z_0$. 

\begin{figure}
	\includegraphics[width=\columnwidth]{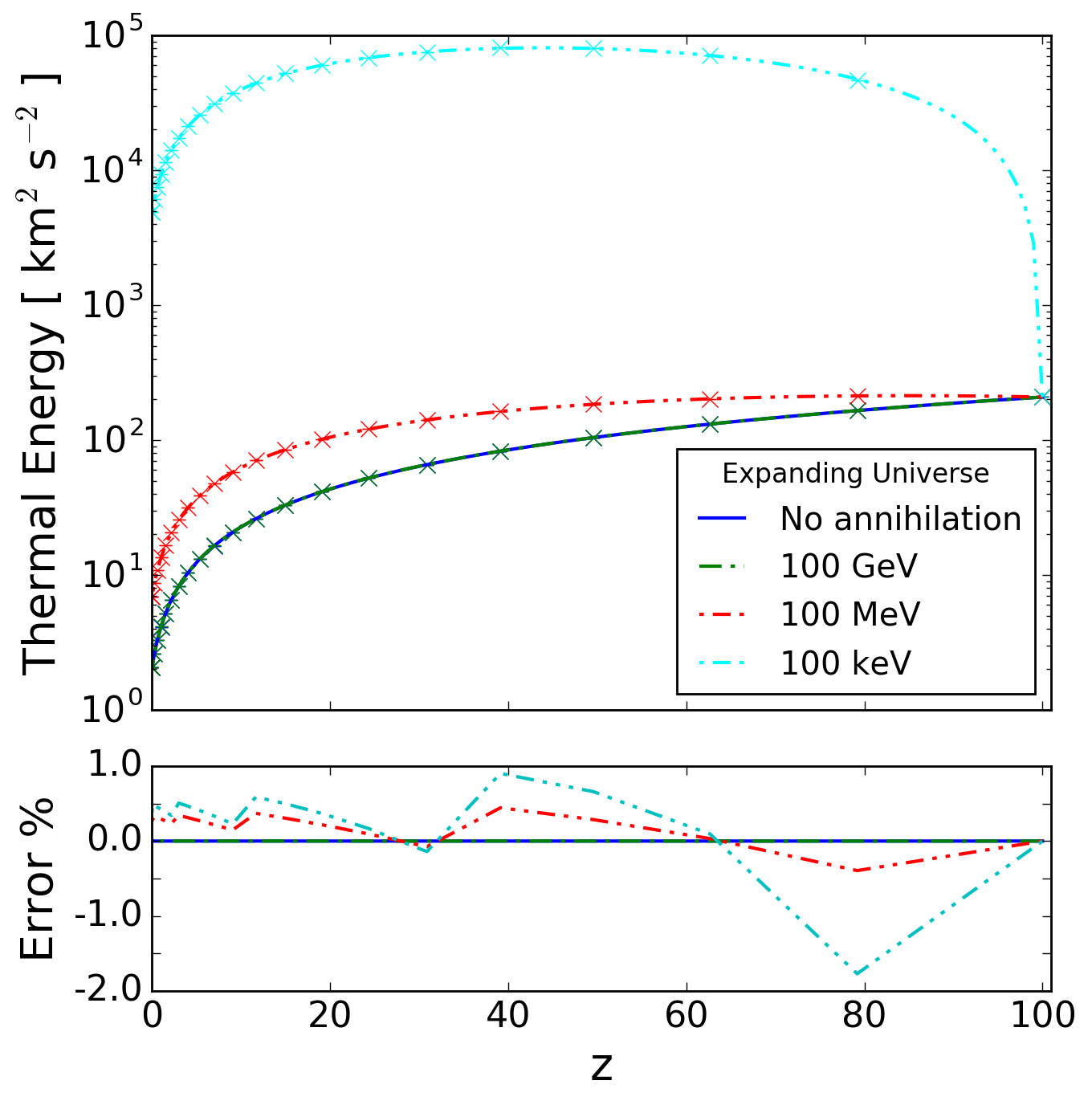}
    \caption{The time evolution of a flat-matter dominated, homogeneous and isotropic universe with an initial gas temperature set to $T=10^{6}$ K and N=16$^3$ particles. Shown in the crosses are the average specific thermal energy from a DMAF simulations with zero perturbations in their initial conditions. The curves are the theoretical expectations as given by equation \ref{eq:Flat_matter_dominated_friedman_Final_equation}. The sub-panel shows the percentage deviation of the simulation results from the expected theoretical curves.}
    \label{fig:dmaf_theory}
\end{figure}

\begin{figure}
	\includegraphics[width=\columnwidth]{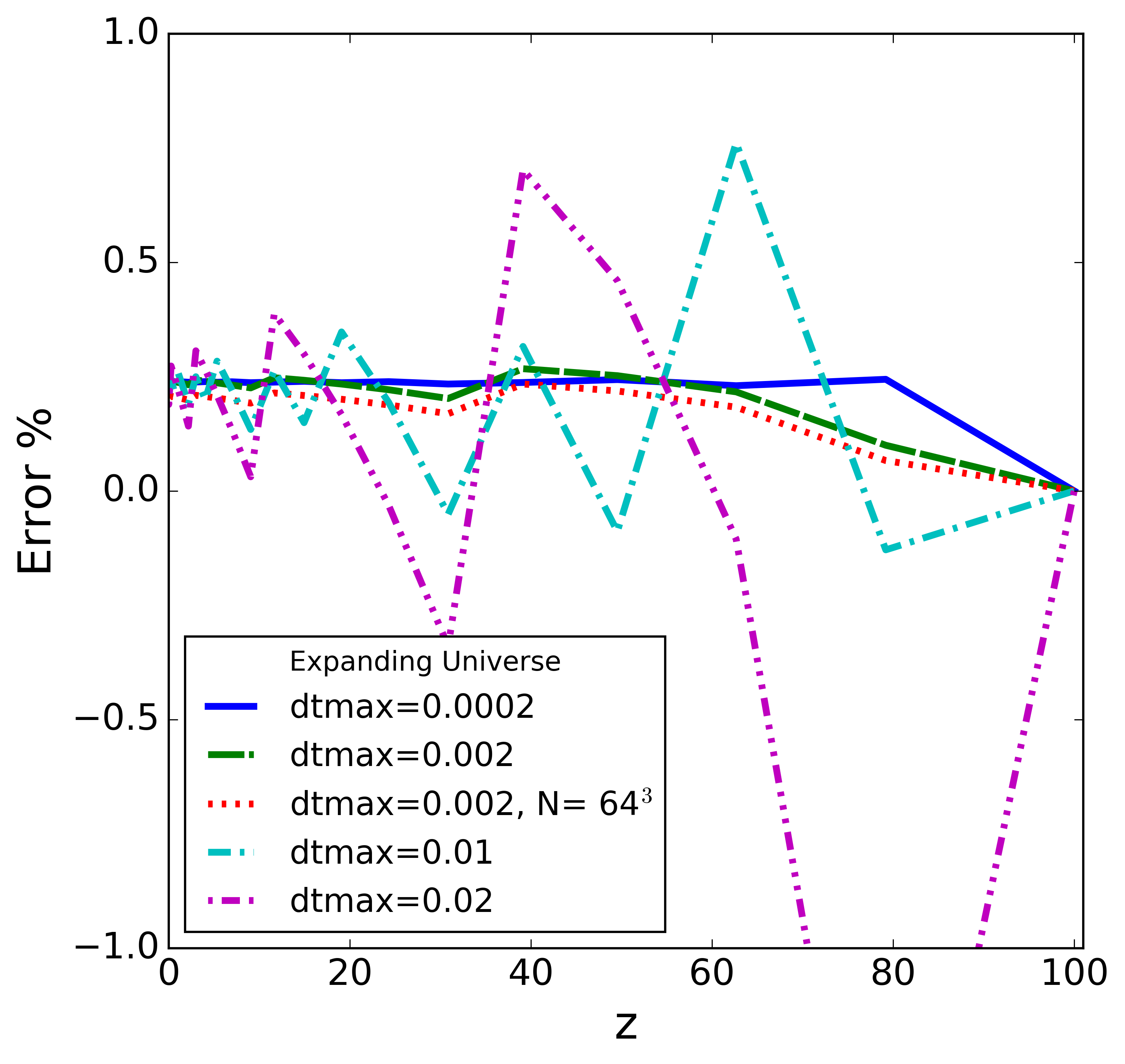}
    \caption{Focusing on the 100 MeV model we show how the uniform simulations behave as the time steps are decreased from as starting temperature of 10 K, to show the error due to the injected energy. We find the error diminishes as the maximal time-steps of the simulation is decreased, converging to an error of < 0.3 percent for dtmax 0.002. This residual error can then be further decreased by increasing particle resolution.}
    \label{fig:dmaf_error}
\end{figure}

Figure \ref{fig:dmaf_theory}, shows the results of an expanding simulation with various DMAF models and the sub panel shows the errors. For the most extreme model we see an error no greater than 2.0 percent (static box simulations yield similar results) which fluctuate about the expected energy. These errors can be minimised by decreasing the maximum allowed time-step of the simulation and increasing the resolution of the simulation as shown in figure \ref{fig:dmaf_error}, demonstrating the robustness of our method.

\section{DMAF uncertainty for cuspy halos}
\label{App:DMAF_halo_error}
The energy density  injected at a point is similar to Equation (\ref{eq:gas_injection})
\begin{equation}
   \frac{de}{dt} = \kappa\rho_{\chi}^{2},
	\label{eq:energy_density_injection}
\end{equation}
where we have multiplied out $\rho_{gas}$ from (\ref{eq:gas_injection}) and $\kappa$ are the appropriate pre-factors.
Focusing on the energy injected within the scale radius of a generic cuspy halo, the density profile is approximately
\begin{equation}
    \rho_{\chi} \propto r^{-\gamma},
	\label{eq:gamma_slope}
\end{equation}
where for $\gamma = 1$ we retrieve the cusp of an NFW profile.
The energy being injected within a volume of radius $r$ in the halo is

\begin{equation}
    \frac{dE(r)}{dt} = 4\pi\kappa\int_{0}^{r} \rho_{\chi}^{2} r'^2 dr' 
	\label{eq:total energy injection}
\end{equation}
For sufficiently shallow cusps with $\gamma < 1.5$, this integral converges and is proportional to 
\begin{equation}
    \frac{dE(r)}{dt} \propto r^{(3-2\gamma)}.
	\label{eq:integration}
\end{equation}
Neglecting DMAF from within an gravitationally unresolved cuspy region at is therefore at most

\begin{equation}
    \Delta = \left(\frac{r_{core}}{r_{s}}\right)^{(3-2\gamma)},
	\label{eq:error}
\end{equation}
where $r_{s}$ is the scale radius and $r_{core}$ is the radius up to which the gravitational force and hence the halo is said to be well resolved i.e near the gravitational smoothing length as argued by \cite{2003Power}. For for $\gamma =1$ we retrieve the approximate uncertainty due to an NFW profile, $\Delta_{nfw} = \frac{r_{core}}{r_{s}}$.


\bsp	
\label{lastpage}
\end{document}